\documentclass{moriond}

\def\be{\begin{equation}}
\def\ee{\end{equation}}
\def\bea{\begin{eqnarray}}
\def\eea{\end{eqnarray}}

\newcommand{\Photo}{}

\begin{document}
\vspace*{4cm}
\title{A gravity antenna based on quantum technologies: MIGA}

\author{B. Canuel$^1$, X. Zou$^1$, D. O. Sabulsky$^1$, J. Junca$^1$, A. Bertoldi$^1$, Q. Beaufils$^2$, R. Geiger$^2$, A. Landragin$^2$,  M. Prevedelli$^3$, S.~Gaffet$^{4}$, D.~Boyer$^{4}$, I.~L{\'a}zaro~Roche$^{4}$, P. Bouyer$^1$}
\address{$^1$LP2N, Laboratoire Photonique, Num{\'e}rique et Nanosciences, Universit{\'e} Bordeaux--IOGS--CNRS:UMR 5298, 1 rue F. Mitterrand, F--33400 Talence, France}
\address{$^2$LNE--SYRTE, Observatoire de Paris, Universit{\'e} PSL, CNRS, Sorbonne Universit{\'e}, 61, avenue de l'Observatoire, F--75014 Paris, France}
\address{$^3$Dipartimento di Fisica e Astronomia, Universit\`{a} di Bologna, Via Berti-Pichat 6/2, I-40126 Bologna, Italy}
\address{$^{4}$LSBB, Laboratoire Souterrain à Bas Bruit,  CNRS UAR3538, Avignon University - La grande combe, 84400 Rustrel, France}

\maketitle\abstracts{
We report the realization of a large scale gravity antenna based on matter-wave interferometry, the MIGA project. This experiment consists in an array of cold Rb sources correlated by a 150 m long optical cavity. MIGA is in construction at the LSBB underground laboratory, a site that benefits from a low background noise and is an ideal premise to carry out precision gravity measurements. The MIGA facility will be a demonstrator for a new generation of GW detector based on atom interferometry that could open the infrasound window for the observation of GWs. We describe here the status of the instrument construction, focusing on the infrastructure works at LSBB and the realization of the vacuum vessel of the antenna.
}

\section{MIGA, an underground array of Atom Interferometers}\label{MIGA_intrument}
\begin{figure}[htp]
  \centering
  \includegraphics[width=0.8\textwidth]{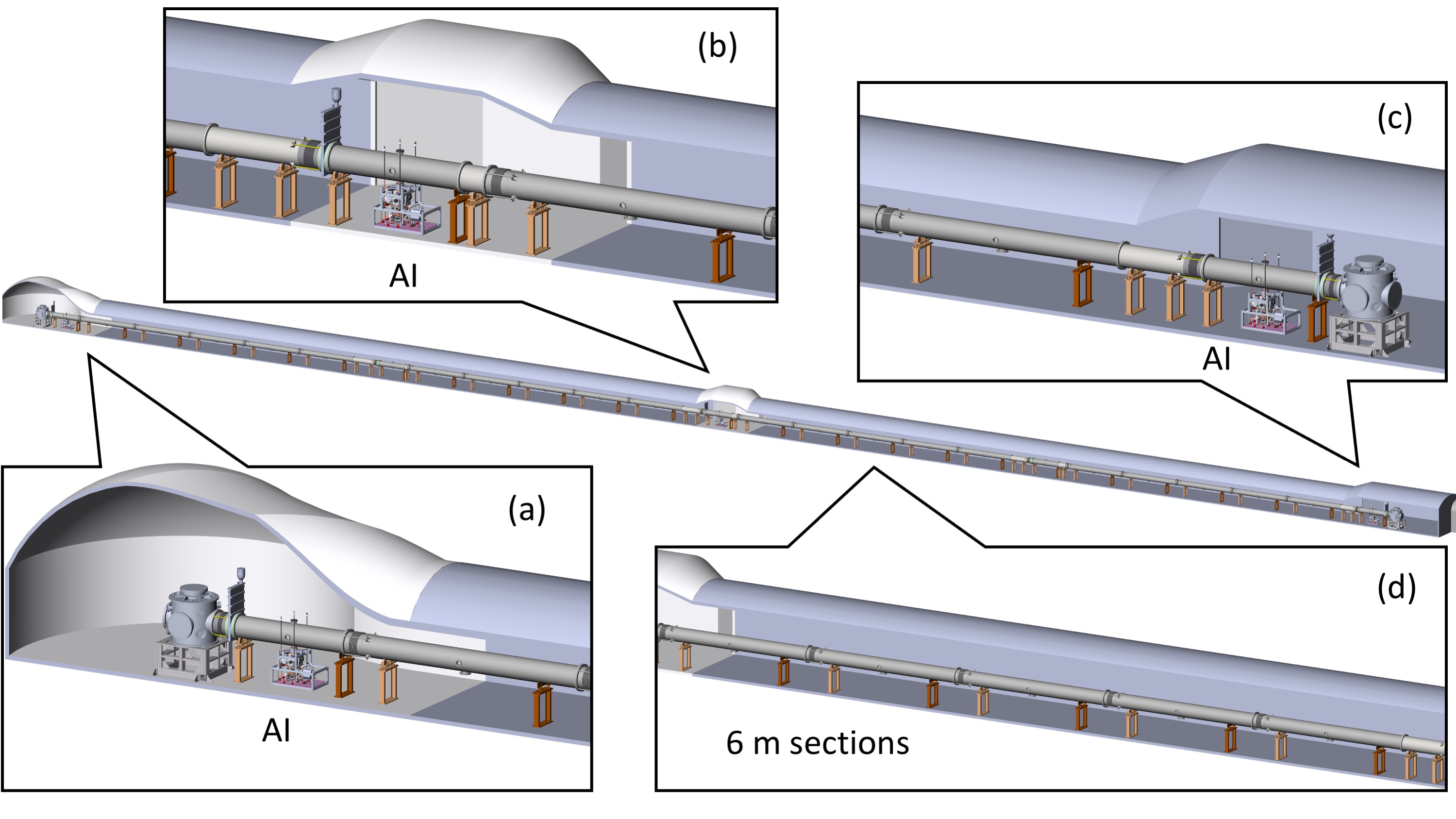}
  \caption{The MIGA instrument. The atom interferometers (AI) of the antenna will be located at (a), (b) and (c). Most of the vacuum vessel is composed of 6 meters long sections (d). }\label{MIGA_antenna}
\end{figure}

MIGA~\cite{CanuelSciRep} consists in a 150 m vacuum vessel hosting in-cavity beams that interrogate and correlate three Rb atom interferometers regularly spaced along the antenna baseline (see Fig.~\ref{MIGA_antenna}).
The vacuum system is mostly composed of standard 6 m long and 50 cm large sections interconnected with metallic gaskets. Two large vacuum tanks are placed at the extremities of the system to accommodate the input optics, the in-cavity mirrors and the control systems of the resonator. 
The atom sources of the experiment~\cite{Beaufils2022} are based on a standard combination of 2D-3D Magneto Optical Traps to produce clouds of Rb atoms at a temperature of $\approx 2$ $\mu$K. After launching the atoms on a vertical trajectory, a set of Raman beams prepares the sources in the pure quantum state $F=2$, $m_F=0$ with a reduced effective temperature of 150 nK before entering the intereferometric region located at their apex where they are manipulated in the Bragg regime using in-cavity fields. All the optical functions of the source for preparation and detection are realized using dedicated fiber laser systems. These commercial solution were specifically developed~\cite{Sabulsky2020} for MIGA to cope for the rough functioning conditions of LSBB in terms of accessibility and environment. The project foresees the production of a total of five Rb sources, three of them dedicated to the antenna and two others for a 6 m demonstrator, a reduced size version of MIGA, developed at the LP2N laboratory. As of today, a total of three systems were produced and fully characterized. In the following, we present the status of the construction of the antenna and detail in particular the infrastructure works recently carried out at the LSBB laboratory to host the experiment and the realization of its vacuum vessel.

\section{Infrastructure works at the MIGA installation site}

The LSBB laboratory comes for the reconvertion of an  underground military command center for ballistic missiles into a low noise research center~\cite{LSBB}. This facility benefits from an exceptional low seimic and magnetic background noise that are major sources of disturbance for precision measurement experiments based on atom interferometry. The installation of MIGA at LSBB required to create new galleries in this environment. Starting at the end of 2018 and till the beginning of 2020, two new perpendicular galleries of 150 m were bored to host the initial antenna and a possible evolution towards a 2D instrument geometry. Indeed, a two-arm, Michelson geometry would present interesting noise rejection properties in view of GW detection to limit the impact of the residual frequency noise of the interrogation laser. These new galleries were realized deep inside the kastic mountain hosting the LSBB: their accesses are located at about 1 km from the entry of the lab and the minimal depth of the MIGA gallery is about 300 m.
\begin{figure}[htp]
  \centering
  \includegraphics[width=1\textwidth]{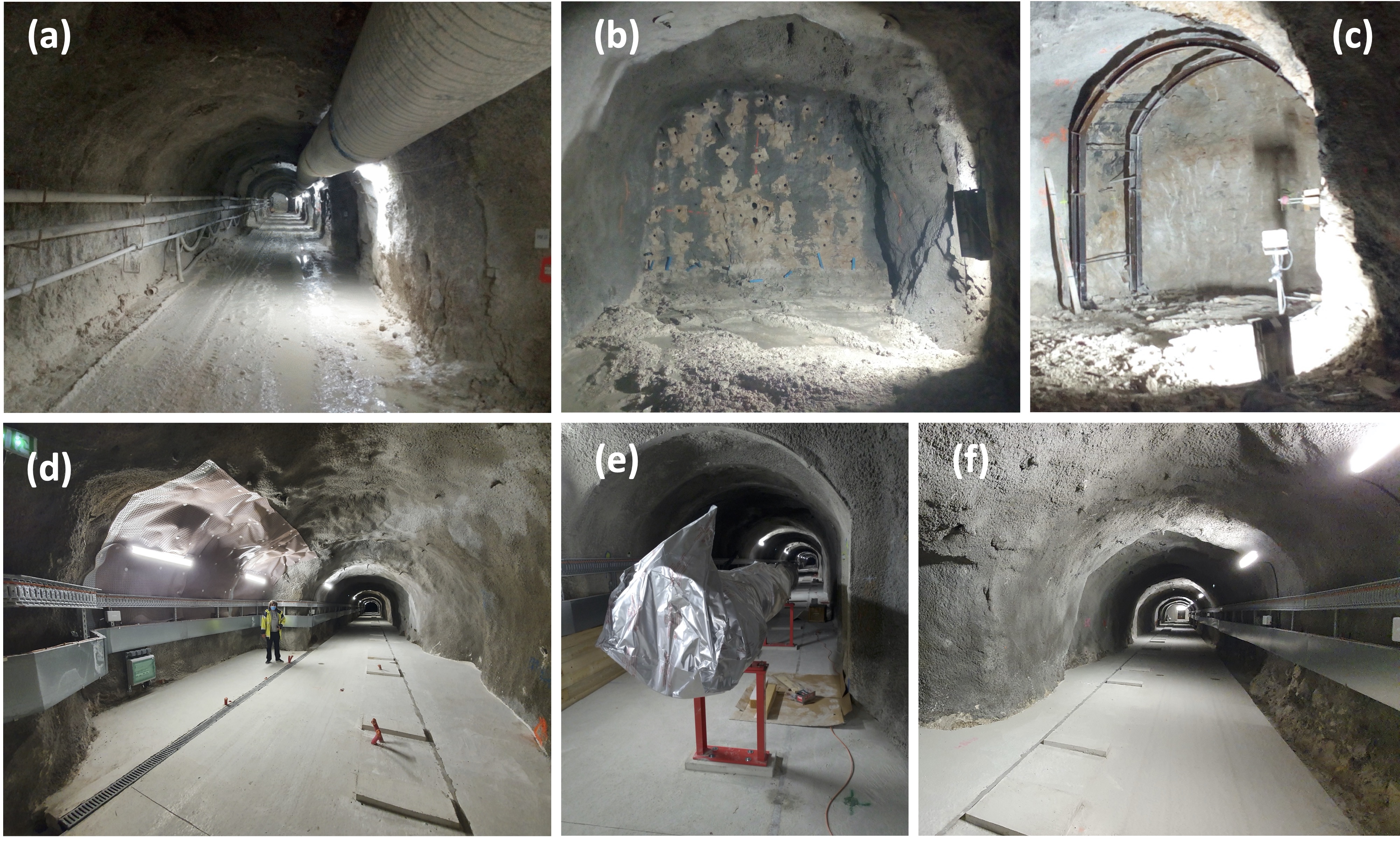}
  \caption{(a)-(b)-(c): the MIGA galleries during the works, (d)-(f): the two finished galleries, (e) installation of the first section of the vacuum vessel within one of the MIGA galleries.}\label{LSBB_MIGA}
\end{figure}

Fig.~\ref{LSBB_MIGA} (a)-(b)-(c)
where taken during the infrastructure works, a period in which the laboratory was closed to any other activity. The works were carried out using explosives, a technique that limited the progress to a few meters per day. Fig.~\ref{LSBB_MIGA} (b) shows a typical firing plan with the position of the different holes for the charges. These calculated positions together with a specific timing for the ignition enabled controlled explosions. Fig.~\ref{LSBB_MIGA} (a) shows the temporary venting system installed to evacuate the dust after firing. After evacuation of the wastes of the explosion, the new portion portion of gallery was reinforced using shotcrete. On some specific locations, the geologic conditions also imposed to install further consolidation using metallic structures (see Fig.~\ref{LSBB_MIGA} (c)).
After the conclusion of the works, a particular care was given to the cleaning of the galleries in order to prepare for the mounting of the UHV system of the antenna. Fig.~\ref{LSBB_MIGA} (d)-(e)-(f) shows the galleries after completion of the works. 
The current gallery is 3 m high and 3,3 m large to accommodate for the installation of the different elements of the vacuum vessel and a lateral pathway for the transportation of the material (see on Fig.~\ref{LSBB_MIGA} (e) the mounting of the first 6 m section of the vacuum system). Larger cavities where also realized to host the end-towers and the atom interferometers: Fig.~\ref{LSBB_MIGA} (d) shows the enlargement done for the installation of the different systems of the central atom interferometer.

\section{MIGA vacuum technology}\label{vacuum_tests}
\begin{figure}[t]
  \centering
   \includegraphics[width=1\textwidth]{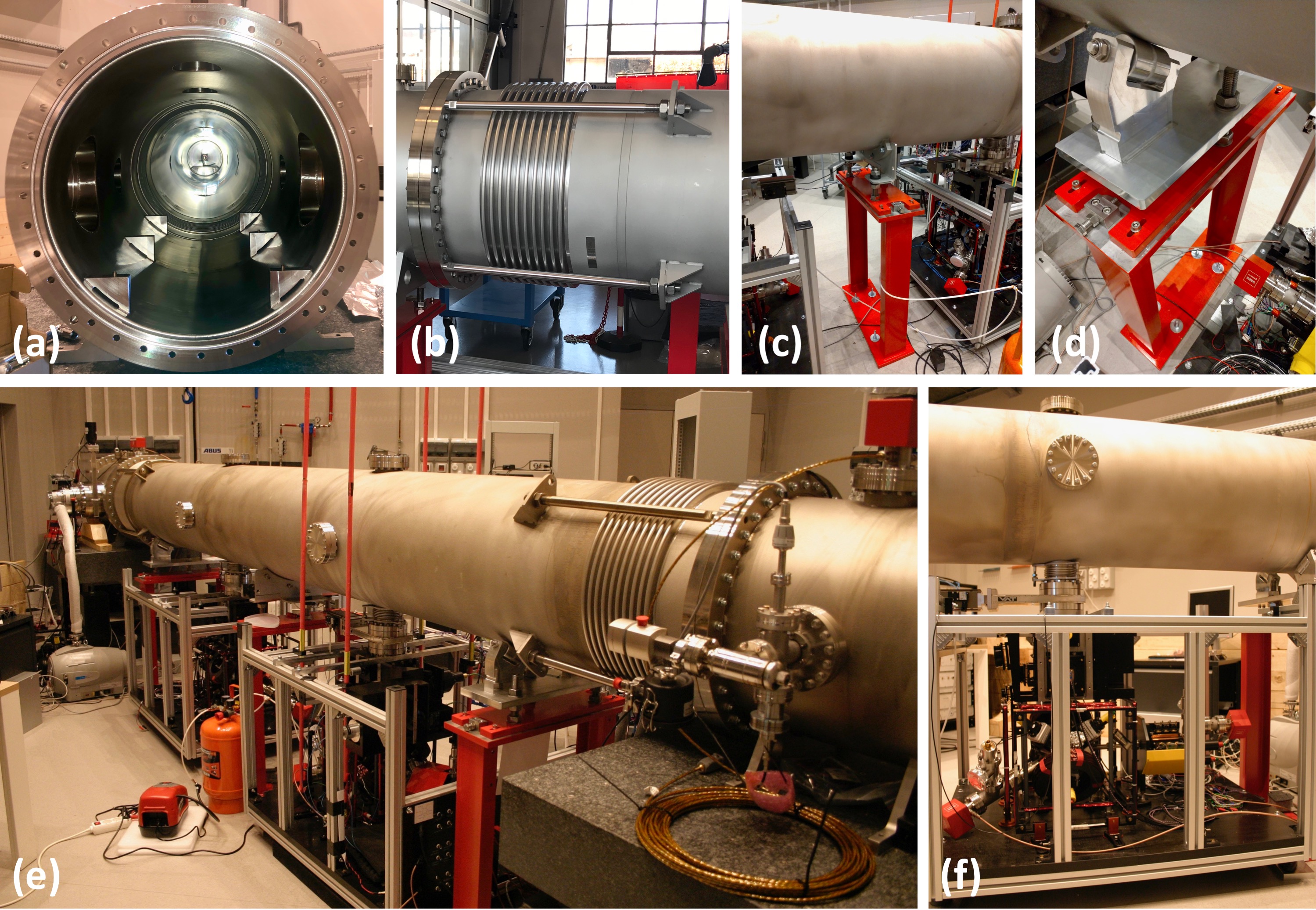}
  \caption{Components of the MIGA vacuum system: (a) 500~mm flanges with ``helicoflex" metallic gaskets, (b) bellows allowing for thermal elongation of the sections, (c)-(d) rigid and flexible supports, (e) prototype atom gradiometer, (f) Rb atom source connected to the gradiometer prototype.}\label{sys_vide}
\end{figure}

The realization of the vacuum system of the antenna is a real technical challenge: the functioning of the atom sources sets a stringent vacuum level in their vicinity of about $10^{-9}$~mbar to avoid atom losses by collision with the residual gas. The total interferometric time of $2T=500$~ms, corresponding to a distance between the cavity interrogation beams of $\approx 30$~cm, imposes the use of a large aperture vacuum system of about 50~cm. The conduction resulting from such a large aperture, imposes a homogeneous vacuum level of $10^{-9}$~mbar in the whole vessel. To reach such performance, the system was designed to allow for a baking procedure of a few days at a maximum temperature of 220°C.
The whole system, including the parts for the 150 m long tube, their supports and the end towers, was produced and tested by SAES Rial Vacuum during year 2020. Another system, based on the exact same technology, was also delivered to assemble a 6,4~m long atom gradiometer that is used as a test bench of the antenna (See Fig.~\ref{sys_vide}).

The vacuum vessel of MIGA is mainly composed of 6~m long, 5~mm thick, SS~304 section which are interconnected using 500~mm large ``helicoflex" metallic gaskets (see Fig.~\ref{sys_vide} (a)). Each section is equipped with a 20~cm long bellow to allow for its thermal elongation during the baking stage (see Fig.~\ref{sys_vide} (b)). The system is placed on different supports allowing for a fine-alignement: some of them enable for a rigid mounting with respect to the ground (see Fig.~\ref{sys_vide} (c)) and some other allow for a free displacement along the antenna direction (see Fig.~\ref{sys_vide} (d)), a flexibility required to ensure a correct fixation during baking. 

The functioning of the system, including the baking and pumping protocol was validated on the prototype gradiometer assembled at LP2N during year 2021 (see Fig.~\ref{sys_vide} (e)-(f)). It is composed of a 4,7~m long section connecting two Rb Atom sources (see Fig.~\ref{sys_vide} (f)). Two 0.8~m long vacuum tanks are placed at the extremities of the system to host the optical setups for the interrogation lasers correlating the atom sources of the gradiometers. The different parts of the vessel were pre-baked at the SAES factory during 24~h at about 220°C before packaging in sealed bags filled with dry nitrogen. After shipping and installation, a set of vacuum tests were carried out at LP2N. After a baking stage to about 100°C for about two days, we could obtain a residual pressure of $8\times10^{-10}$~mbar, corresponding to an outgassing rate of $6.3\times10^{-12}$~mbar$\cdot$l$\cdot$s$^{-1}\cdot$cm$^{-2}$. If the same performances are obtained with the vacuum components of the antenna, a total pumping speed of 10000~l$\,\cdot\,$s$^{-1}$ will be sufficient to obtain a residual pressure better than $2\times10^{-9}$~mbar.

\section{Conclusion}\label{}
We reported here the major steps achieved towards the realization of MIGA: the infrastructure works required to install the antenna were successful; three of the Rb atom sources are now realized and tested; the vacuum system has been fully produced and the baking and pumping protocol validated. The assembly of the system is now starting at LSBB and the connection of the two first atom sources is expected during year 2022. After completion, MIGA will be the first large scale experiment based on matter wave interferometry and will pave-the-way towards research infrastructures that could study gravity with unprecedented precision, like ELGAR~\cite{ELGAR2020}.


%
%
%
%
%
%

\end{document}